\documentclass[11pt,a4paper]{article}
\usepackage{jstyle}
\usepackage{slashed}

\newcommand{\bi}{\begin{itemize}}
\newcommand{\ei}{\end{itemize}}
\newcommand{\bea}{\begin{eqnarray}}
\newcommand{\eea}{\end{eqnarray}}

\author{Euihun JOUNG}
\author{\qquad Wenliang LI}

\affiliation{Department of Physics and Research Institute of Basic Science, Kyung Hee University, Seoul 02447, Korea}

\emailAdd{euihun.joung@khu.ac.kr}
\emailAdd{lii.wenliang@gmail.com}

\title{\centering
\LARGE{
Non-Relativistic Limits of \\ Colored Gravity in Three Dimensions
}}

\abstract{
The three-dimensional non-relativistic isometry algebras, namely Galilei and Newton-Hooke algebras,  
are known to admit double central extensions,
which allows for non-degenerate bilinear forms hence for action principles through Chern-Simons formulation. 
In  three-dimensional colored gravity,
the same central extension  helps the theory evade the multi-graviton no-go theorems by enlarging the color-decorated isometry algebra. 
We investigate the non-relativistic limits of three-dimensional colored gravity in terms of generalized \.In\"on\"u-Wigner contractions. 
}

\begin{document}


\maketitle

\section{Introduction}\label{sec-intro}
In the non-relativistic limit, Einstein's general relativity reduces to Newtonian gravity. 
In analogy to Einstein-Cartan gravity, Newtonian gravity has a coordinate-independent formulation in terms of frame fields, namely Newton-Cartan  gravity \cite{Cartan:1923zea,Cartan:1924yea,Duval:1984cj,Bekaert:2014bwa, Bekaert:2015xua}. 
Since Einstein-Cartan gravity can be obtained by gauging the relativistic Poincar\'e algebra, 
one may attempt to derive Newton-Cartan gravity by gauging a non-relativistic symmetry algebra. 
The first candidate algebra would be the Galilei algebra ${\mathfrak g}_0$, which is the non-relativistic limit of the Poinca\'re algebra. 
However, it turns out that one instead needs to consider the Bargmann algebra ${\mathfrak g}_{0}^+$, 
which is an extension of the Galilei algebra with a central generator $M$ called ``mass''. 
This mass generator is related to an additional $U(1)$ gauge field in Newton-Cartan gravity. 
If we introduce a non-zero cosmological constant, 
then the counterparts of the Galilei algebra ${\mathfrak g}_0$ and the Bargmann algebra ${\mathfrak g}_0^+$
are the Newton-Hooke algebra ${\mathfrak g}_\L$ and the extended Newton-Hooke algebra ${\mathfrak g}_\L^+$. 

Since the equations of motion of Newton-Cartan gravity 
can be obtained by gauging the Bargmann algebra ${\mathfrak g}_{0}^+$ \cite{Andringa:2010it}, 
Newton-Cartan gravity can be considered as a gauge theory of the Bargmann algebra. 
Another way to derive Newton-Cartan gravity is to consider a limit of Einstein-Cartan gravity 
mimicing the \.In\"on\"u-Wigner contraction of the relativistic algebra \cite{Bergshoeff:2015uaa}. 
However, in general spacetime dimensions, these equations of motion do not have a simple action principle.

In three dimensions, the Bargmann algebra ${\mathfrak g}_0$ admits a second central extension by the ``spin'' generator $S$ \cite{Grigore:1993fz, Jackiw:2000tz}. 
The resulting algebra is referred to as the extended Bargmann algebra ${\mathfrak g}_0^{++}$\,.
In parallel, in the cosmological case, the three-dimensional Newton-Hooke algebra admits double extensions.
We refer to this as the doubly extended Newton-Hooke (NH) algebra ${\mathfrak g}_{\L}^{++}$\,.
We summarize the zoo of three-dimensional non-relativistic algebras in Table \ref{3d-nonrel-alg}. 
Note that the subscript $\L$ denotes the cosmological constant and the superscript $+$ or $++$ means the central extension
by $M$ or $M, S$\,.
Due to the second central extension, 
both ${\mathfrak g}_0^{++}$ and ${\mathfrak g}_{\L}^{++}$ have non-degenerate invariant bilinear forms \cite{Papageorgiou:2009zc, Papageorgiou:2010ud}.
Consequently, the corresponding theories admit action principles through  Chern-Simons formulation.
These  non-relativistic gravity theories are sometimes called extended Bargmann gravity and extended Newton-Hooke (NH) gravity, respectively.
Along the same lines, the non-relativistic Chern-Simons theories admit various extensions according to the symmetry algebras 
\cite{Hartong:2016yrf, Bergshoeff:2016lwr,Bergshoeff:2016soe}. 

\begin{table}[h!]
\centering
 \begin{tabular}{|c || c | c |} 
 \hline
  & Flat & (A)dS \\
 \hline
 No extension & Galilei  ${\mathfrak g}_0$ & Newton-Hooke (NH) ${\mathfrak g}_\L$ \\
 Extended by $M$  & Bargmann ${\mathfrak g}_{0}^+$ & Extended NH ${\mathfrak g}_\L^+$ \\
 Extended by $M$, $S$ &  Extended Bargmann  ${\mathfrak g}_{0}^{++} $ &
  Doubly extended NH ${\mathfrak g}_\L^{++}$ \\
 \hline
\end{tabular}
\caption{Three-dimensional non-relativistic isometry algebras.}
\label{3d-nonrel-alg}
\end{table}

The presence of two central generators $M$ and $S$ are hence crucial to 
the action principle of three-dimensional non-relativistic gravities in the frame formulation.
In fact, as we shall show later, these doubly extended non-relativistic algebras
can be obtained from the relativistic ones with two $u(1)$ generators after suitable contractions (see also \cite{Bergshoeff:2016lwr, Hartong:2017bwq}). 
For instance, in AdS, the relativistic isometry algebra with two $u(1)$'s,
\be
	so(2,2)\oplus u(1) \oplus u(1)
	= \Big(sl(2,\mathbb R)\oplus u(1)\Big)\oplus \Big(sl(2,\mathbb R) \oplus u(1)\Big)\,,
	\label{assoc}
\ee
contracts to the doubly extended NH ${\mathfrak g}_\L^{++}$ under a suitable
scaling limit. In fact, both the relativistic and non-relativistic algebras
can be split into left- and right-moving sectors, so the contraction can be performed in each sector.  
After the contraction, the $u(1)$ generators become central elements,
and the resulting algebra cannot be arranged as a direct sum anymore.
The is the key mechanism for a non-degenerate bilinear form --- hence for the action principle.

We notice that another peculiar utility of $u(1)$ can be found in a different extension of 
three-dimensional gravity, namely the colored gravity
which has been studied recently in \cite{Gwak:2015vfb,Gwak:2015jdo, Joung:2017hsi}. 
Several no-go results are known for theories of multiple gravitons,\footnote{See e.g. \cite{Boulanger:2000rq} for the no-go result and \cite{Boulanger:2000ni} for an exception.} 
but in three dimensions 
they can be evaded by extending the isometry algebra $so(2,2)$ with 
two $u(1)$ generators.
It is remarkable that the resolution of the problems in non-relativistic limit and color-decoration
are the same: replace $sl(2,\mathbb R)$ by $sl(2,\mathbb R)\oplus u(1)$. 
This observation leads us to the following question: 
does the three-dimensional colored gravity have a natural non-relativistic limit? 

Let us explain a few more details about the colored gravity.
It can be constructed in the Chern-Simons formulation 
with an appropriate choice of gauge algebra.
For the multi-graviton interpretation, this algebra should contain 
multiple copies of isometry generators,
hence a direct product of the isometry $so(2,2)$ and 
the internal symmetry.
Usually, the direct product of two Lie algebras does not define a Lie algebra, 
but the case of two associative algebras does so. 
The internal symmetry can be taken as $U(N)$ which obviously has an associative structure.
About the isometry algebra, we can extend the Lie algebra
$sl(2,\mathbb R)\oplus sl(2,\mathbb R)$ to
an associative algebra
  $gl(2,\mathbb R)\oplus gl(2,\mathbb R)$
  by adding  two copies of $u(1)$.
 We refer to the internal symmetry as ``color symmetry'' for the reason which will become 
clear later, 
and the resulting gravity theory will be correspondingly referred to as ``colored gravity''.
The colored gravity has several interesting features:
{\it i)} It has a non-trivial potential with many saddles points 
corresponding to the (A)dS vacua of different cosmological constants.
These vacua can be classified according to how much the internal symmetry is preserved. 
{\it ii)} Around the color singlet vacuum, the theory has one graviton, $(N^2-1)$ massless spin-two fields charged under the color symmetry 
and $U(N)\times U(N)$ Chern-Simons gauge field.
The color non-singlet spin-two fields are strongly interacting for large $N$.
{\it iii)} Around a color-symmetry breaking background,
a Higgs-like mechanism takes place and one finds partially-massless spin-two fields 
\cite{Deser:1983tm,Deser:2001us}
appearing in the spectrum.

The organization of the paper is as follows. 
In Section \ref{sec-non-relativistic}, 
we present a concise overview of several aspects of the 3d non-relativistic algebras. 
In Section \ref{sec-non-relativistic-colored-alg}, 
we derive the non-relativistic colored-gravity algebras 
from the \.In\"on\"u-Wigner contractions of the relativistic ones. 
In Section \ref{sec-non-relativistic-colored-action}, 
we construct the actions of non-relativistic colored gravities using the Chern-Simons formulation. 
In Section \ref{sec: discussion}
we conclude the work with a brief summary. 
Appendix includes technical details and additional materials.


\section{Non-relativistic Algebras and their Bilinear Forms}\label{sec-non-relativistic}

\subsection{Various non-relativistic algebras in three dimensions}\label{subsec-3d-nonrel}

We will give a concise review of various three-dimensional non-relativistic algebras: 
the Galilei, Bargmann, extended Bargmann, Newton-Hooke (NH), extended NH, and doubly extended NH algebras, 
which have been briefly discussed in Introduction.

\subsubsection*{Galilei algebra ${\mathfrak g}_0$}

The three-dimensional Galilei algebra  is an eight-dimensional Lie algebra with the non-trivial Lie brackets:
\be 
[J,\,G_i]=2\,\epsilon_{ij}\, G_j,\qquad
[J,\,P_i]=2\,\epsilon_{ij}\,  P_j, \qquad
[H,\,G_i]=2\,\epsilon_{ij}\, P_j.\label{Gal}
\ee
Here, $H, P_i, J$ and $G_i$ are the generators of 
time translations, spatial translations, spatial rotations and Galilean boosts, respectively.
The Levi-civita symbol $\e_{ij}$ is defined as
\be
\epsilon_{12}=-\epsilon_{21}=1,
\quad\epsilon_{11}=\epsilon_{22}=0.
\ee
Remark that, in higher dimensions, there exist non-zero brackets between two spatial rotation generators. 

\subsubsection*{Newton-Hooke algebra ${\mathfrak g}_\L$}

When the cosmological constant $\L$ is turned on,
the Galilei algebra is deformed to the so-called Newton-Hooke algebra.
As a result, we have an additional set of Lie brackets, 
\be
[H,\,P_i]=-2\,\Lambda\,\epsilon_{ij}\, G_j, \label{NH} 
\ee
besides those in \eqref{Gal}.

\subsubsection*{Bargmann algebra ${\mathfrak g}_0^+$ and extended NH algebra ${\mathfrak g}_\L^+$}

In general dimensions, both the Galilean and Newton-Hooke (NH) algebras can be centrally extended by the mass generator $M$.\footnote{In a relativistic theory, 
the energy of a particle in the rest frame  
is proportional to the rest mass. 
However, in a non-relativistic theory, 
spacetime symmetries do not require energy-mass equivalence, 
so we can consider an independent generator for mass 
in the form of a central extension. 
Intuitively speaking, the energy-mass relation is lost in the large speed of light limit.
}
The resulting algebras are known as the Bargmann algebra and the extended NH algebra, respectively.
In three dimensions, the additional Lie brackets are
\be
[G_i, P_j]=2\,\epsilon_{ij}\,M.\label{Bar}
\ee

\subsubsection*{Extended Bargmann algebra ${\mathfrak g}_0^{++}$ 
and doubly extended NH algebra ${\mathfrak g}_\L^{++}$}

In three dimensions, the Bargmann and extended NH algebras admit a second central extension by 
the spin generators $S$.\footnote{To interpret $S$ as a spin generator is subtle.
See \cite{Duval:2002cw, Hagen:2002pg} for more details.}
The extended Bargmann algebra, or equivalently the doubly extended Galilei algebra, has one more set of 
Lie brackets, 
\be
	[G_i,G_j]=2\,\epsilon_{ij}\, S.\label{ext}
\ee
In the doubly extended Newton-Hooke algebra, 
the additional Lie brackets are \eqref{ext} and
\be
[P_i,P_j]=-2\,\Lambda\,\epsilon_{ij}\,S.\label{extNH}
\ee

\medskip

All the non-relativistic algebras presented above can be
obtained from the most general one, the doubly extended NH algebra, by various \.In\"on\"u-Wigner contractions.
The cosmological constant dependent terms can be removed by the usual \.In\"on\"u-Wigner contraction.
The mass and spin generators only appear in the right hand sides of the Lie brackets,
hence can be eliminated by a suitable contraction:  rescale them first by a contraction parameter, 
 then send the parameter to zero.

\subsection{Non-relativistic limits as contractions}

It is useful to realize these non-relativistic limits as contractions. 
In fact, the three-dimensional relativistic isometry algebra with a non-zero cosmological constant and two additional $u(1)$ generators
can contract to the doubly extended NH algebra (hence all non-relativistic algebras discussed in section \ref{subsec-3d-nonrel}).
Let us provide more details on this point.
The relativistic isometry algebra in three dimensions has the Lie brackets,
\be
	[\,\mathcal J_{a}\,,\,\mathcal J_{b}\,]=2\,\epsilon_{ab}{}^{c}\,\mathcal J_{c}\,,
	\qquad
	[\,\mathcal J_a\,,\,\mathcal P_{b}\,]=2\,\epsilon_{ab}{}^{c}\,\mathcal P_{c}\,,
	\qquad
	[\,\mathcal P_{a}\,,\,\mathcal P_{b}\,]=-2\,\Lambda\,\epsilon_{ab}{}^{c}\,\mathcal J_{c}\,,
	\label{com rel}
\ee
where $\cP_a$ and $\cJ_a$ are the translation and Lorentz generators,
and
 $\L$ is the cosmological constant.

\subsubsection*{Contraction to Newton-Hooke algebra}

For an illustration, let us see how 
the relativistic isometry algebra \eqref{com rel}
contracts to the Newton-Hooke algebra ${\mathfrak g}_\L$.
We first redefine the generators using the parameter $c$ as
\be
G_i=\frac 1 {c}\,\mathcal J_i\,,
\qquad
P_i=\frac1{c}\,\mathcal P_i\,,
\label{PG}
\ee
and
\be
J=\cJ_0\,,\qquad 
H=\cP_0\,,
\ee
then derive their Lie brackets.
Notice that the temporal components scale differently in $c$ from the spatial ones.
The \.In\"on\"u-Wigner contraction is performed in the $c\to \infty $ limit which 
effectively removes certain terms in the structure constants.
One can easily check that the resulting algebra is the NH algebra ${\mathfrak g}_\L$.
Hence, one can view this contraction  as  the non-relativistic limit 
and interpret the parameter $c$ as the speed of light.

\subsubsection*{Contraction to doubly extended Newton-Hooke algebra}

Let us now show how the 
doubly extended NH algebra $\mathfrak g^{++}_{\L}$ 
can be derived by a contraction.
As the algebra $\mathfrak g^{++}_{\L}$ has dimension eight due to the central generators $M$ and $S$,
it cannot be obtained as a contraction of the relativistic isometry algebra which has dimension six.
To fix the mismatch between the numbers of generators, we need to include two additional $u(1)$ generators.
We refer to these generators as $\cZ_\cP$ and $\cZ_\cJ$.
In terms of them, we change the definitions of $H$ and $J$ to
\be
J=\cZ_\cJ+\cJ_0, \qquad 
H=\cZ_\cP+\cP_0,
\label{HJ mix}
\ee
but keep those of $P_i$ and $G_i$ the same as \eqref{PG}.
The mass and spin generators are given by the other combinations with different rescalings,
\be
	S=\frac {\cZ_\cJ-\cJ_0}{2\,c^2},
	\qquad
	M=\frac {\cZ_\cP-\cP_0}{2\,c^2}.
	\label{MS mix}
\ee
With these definitions, one can succesfully reproduce
the doubly extended NH algebra $\mathfrak g_\L^{++}$
by taking the $c\to \infty$ limit.
Note that in \eqref{HJ mix} and \eqref{MS mix} the mixings with $\cZ_\cP$ and $\cZ_\cJ$ are independent,
hence the corresponding central extensions by $M$ and $S$ are also independent. 

\subsubsection*{Chiral decomposition}

When $\L\neq0$\,, the relativistic isometry algebra can be decomposed into two parts:
\be
	\cL_a=\frac 1 2\left(\cJ_a+\frac1{\sqrt{-\L}}\,\cP_a\right)\,,
	\qquad 
	\tilde\cL_a=\frac 1 2\left(\cJ_a-\frac1{\sqrt{-\L}}\,\cP_a\right)\,,
	\label{LL JP}
\ee
and each of them forms an $sl(2)$\,:
\be
	[\cL_a,\cL_b]=2\,\e_{ab}{}^c\,\cL_c\,,
	\qquad
	[\tilde\cL_a,\tilde\cL_b]=2\,\e_{ab}{}^c\,\tilde\cL_c\,.
\ee
Accordingly, it is natural to decompose the $u(1)\oplus u(1)$ part as 
\be
	\cZ=\frac 1 2\left(\cZ_\cJ+\frac 1 {\sqrt{-\L}}\,\cZ_\cP\right),\qquad 
	\tilde\cZ=\frac 1 2\left(\cZ_\cJ-\frac 1 {\sqrt{-\L}}\,\cZ_\cP\right).
	\label{chiral-Z}
\ee

For the non-relativistic limit $c\to \infty$, 
one can check that the redefinitions \eqref{PG}, \eqref{HJ mix} and \eqref{MS mix}
are compatible with the chiral decomposition.
More precisely, with the redefinitions,
\be
	K=\cZ+\cL_0,\qquad
	L_i=\frac1c\,\cL_i,
	\qquad
	Z=\frac{\cZ-\cL_0}{2\,c^2},
	\label{chiral def}
\ee
\be
	\tilde K=\tilde \cZ+\tilde \cL_0,\qquad
	\tilde L_i=\frac1c\,\tilde \cL_i,
	\qquad
	\tilde Z=\frac{\tilde \cZ-\tilde \cL_0}{2\,c^2},
	\label{anti-chiral def}
\ee
the two copies of the relativistic chiral algebra $sl(2)\oplus u(1)$ contract to
the following non-relativistic chiral algebras,
\be
	[K,\,L_i]=2\,\e_{ij}\,L_j, \qquad
	[L_i,\, L_j]=2\,\e_{ij}\,Z,
\ee
\be
	[\tilde K,\,\tilde L_i]=2\,\e_{ij}\,\tilde L_j,\qquad
	[\tilde L_i,\, \tilde L_j]=2\,\e_{ij}\,\tilde Z,
\ee
where $Z$ and $\tilde Z$ are the centers of the chiral algebras.  
The chiral subalgebra $\mathfrak{h}$, spanned by $K, L_i$ and $Z$, can be viewed as 
a central extension of
the two-dimensional Euclidean algebra
where $K$ and $L_i$ can be interpreted as two-dimensional translation and rotation generators, respectively.
One can equally interpret  $\mathfrak{h}$ as a harmonic oscillator, 
where $L_1\pm i\,L_2$ are creation/annihilation or position/momentum operators, 
$K$ is the Hamiltonian, and $Z$ is the Plank contant $\hbar$.\footnote{The same 
algebra can be obtained from the subalgebra of Virasoro algebra spanned by
$L_m/m^{3/2}$, $L_{-m}/m^{3/2}$ and $L_0/m$ (here $L_m$ are the standard generators of Virasoro algebra)
in the $m\to\infty$ limit.  The Virasoro central charge $c$ becomes the central element $Z$.}

The doubly extended NH  algebra $\mathfrak g_{\L}^{++}$ can be directly decomposed into 
two copies of  the chiral algebra $\mathfrak h$ and $ \tilde {\mathfrak h}$\,:
\be 
	\mathfrak g_{\L}^{++}=\mathfrak h \oplus \tilde {\mathfrak h},
	\label{chiral decomp}
\ee
where the explicit decomposition reads
\ba
	G_i=L_i+\tilde L_i,\qquad 
	&J=K+\tilde K,\qquad 
	&S=Z+\tilde Z,\nn
	P_i={\sqrt{-\L}}\,  (L_i-\tilde L_i),\qquad 
	&H={\sqrt{-\L}}\, (K-\tilde K),\qquad 
	&M={\sqrt{-\L}}\, (Z-\tilde Z).
	\label{DENH decomp}
\ea

\subsection{Invariant bilinear form}
\label{sec: inv bf}

One may attempt to use the non-relativistic algebras in section \ref{subsec-3d-nonrel} to construct 
non-relativistic gravity theories.
In three dimensions, as in the relativistic case, one can rely on the Chern-Simons formulation
where non-degeneracy of the bilinear form is important for 
the dynamics of the resulting theory. 
It turns out that, among the algebras that we have considered,
only those with double extensions by $M$ and $J$ admit a non-degenerate bilinear form.
In below, we focus on the doubly extended NH algebra and 
discuss its bilinear form.

Since the doubly extended NH algebra $\mathfrak g_\L^{++}$ admits
the decomposition \eqref{chiral decomp},
we can consider 
the invariant bilinear form $\la \cdot, \cdot \ra$ for each chiral sector.
The non-diagonal part are the $2\times 2$ symmetric matrices,
\be
	B=\begin{pmatrix}
	\la K, K \ra & \la Z, K\ra \\
	\la K, Z \ra & \la Z, Z \ra
	\end{pmatrix},
	\qquad
	\tilde B=\begin{pmatrix}
	\la \tilde K, \tilde K \ra & \la \tilde Z, \tilde K\ra \\
	\la \tilde K, \tilde Z \ra & \la \tilde Z, \tilde Z \ra
	\end{pmatrix}.
\ee
Due to the rotational symmetry, the diagonal part is fixed as
\be
	\la L_i, L_j \ra=\ell\,\delta_{ij},
	\qquad \la \tilde L_i, \tilde L_j \ra=\tilde \ell\,\delta_{ij}\,,
\ee 
with undetermined constants $\ell$ and $\tilde \ell$,
while the other bilinear forms $\la L_i, K\ra$, $\la L_i, Z\ra$ and  their tilde counterparts vanish.
Symmetry also determines the matrices $B$ and $\tilde B$ to be
\be
	B=\begin{pmatrix}
	\b & \ell\, \\
	\ell & 0
	\end{pmatrix},
	\qquad
	\tilde B=\begin{pmatrix}
	\tilde \b & \tilde \ell \\
	\tilde \ell & 0
	\end{pmatrix},
	\label{bl L}
\ee
where $\b$ and $\tilde \b$ are additional allowed constants. 
The  non-degeneracy of the bilinear form requires  $\ell\neq0$ and $\tilde \ell\neq0$,
but $\b$ and $\tilde \b$ are left unconstrained.

The bilinear form of the doubly extended NH algebra
\eqref{DENH decomp} is given by those of the left and right chiral algebras $\mathfrak h$ and $\tilde{\mathfrak h}$, 
together with possible cross terms. 
However, one can show that the only cross term 
compatible with symmetry is
\be
	\la K,\tilde K\ra=\g.
	\label{K tilde K}
\ee
In the end, the invariant bilinear form of $\mathfrak{g}_\L^{++}$
has  five undetermined constants: 
\be
	\ell,\quad  \tilde \ell,\quad  \b, \quad \tilde \b, \quad \g.
\ee
The number of free parameters is consistent with the analysis of \cite{Hartong:2016yrf}.

As in the usual relativistic Chern-Simons gravity, we set $\ell$ and $\tilde\ell$ as
\be
	\ell=-\tilde \ell=\frac{1}{2\sqrt{-\L}}\,,
\ee
 so the spatial part takes the standard form.
The non-vanishing components of the bilinear form are
\be
	\la J, M\ra=\la S, H\ra=1\,,\qquad
	\la P_i, G_j\ra=\delta_{ij}\,,
\ee
and
\be
	\begin{pmatrix}
	\la H,H \ra & \la H,J\ra \\
	\la J,H \ra & \la J,J \ra
	\end{pmatrix}
	=
	\begin{pmatrix}
	-\L\,(\b+\tilde \b -2\,\g) &{\sqrt{-\L}}\,( \b-\tilde \b)\\
	{\sqrt{-\L}}\,( \b-\tilde \b) &  \b+\tilde \b+2\,\g
	\end{pmatrix}.
\ee
The above symmetric matrix remains completely arbitrary.

The bilinear form that we have derived using symmetry can also be obtained from the relativistic one by a contraction.
To begin with, let us focus on the chiral sector. The bilinear form of $\mathfrak{h}=sl(2)\oplus u(1)$ 
is unique for each of the $sl(2)$ and $u(1)$ part, up to overall factors:
\be
	\la \cZ,\cZ\ra=\z,\qquad
	\la \tilde\cZ,\tilde\cZ\ra=\tilde\z,
	\label{rel bil}
\ee
\be
	\la \cL_a,\cL_b\ra =\l\, \eta_{ab},\qquad
	\la \tilde \cL_a,\tilde \cL_b\ra =\tilde \l\, \eta_{ab}.
	\label{LL-tr}
\ee
Using \eqref{chiral def} and \eqref{anti-chiral def}, 
these give the bilinear form for the rescaled generators as
\be
	\la L_i,L_j\ra =\frac{\l}{c^2}\,\delta_{ij}\,,
	\qquad
	\la \tilde L_i,\tilde L_j\ra =\frac{\tilde \l}{c^2}\,\delta_{ij}\,,
\ee
\be
	B=\begin{pmatrix}
	\z-\l & ({\z+\l})/(2c^2) \\
	({\z+\l})/(2c^2) & ({\z-\l})/(4c^4)
	\end{pmatrix},
	\quad
	\tilde B=\begin{pmatrix}
	\tilde \z-\tilde \l & (\tilde \z+\tilde \l)/(2c^2) \\
	(\tilde \z+\tilde \l)/(2c^2) & (\tilde \z-\tilde \l)/(4c^4)
	\end{pmatrix}.
\ee
Hence by choosing
\be
	\z=c^2\,\ell+ \b/2 ,\quad
	\tilde \z=c^2\,\tilde\ell+\tilde \b/2,\quad
	\l=c^2\,\ell- \b/2,\quad 
	\tilde\l=c^2\,\tilde \ell-{\tilde \b}/2,
	\label{l z}
\ee
we recover the bilinear form \eqref{bl L} in the contraction limit.
Then we consider the brackets of the left and right sectors. 
We have the freedom to introduce the cross term,
\be
	\la \cZ, \tilde \cZ\ra=\g,
\ee
which leads to \eqref{K tilde K} after the contraction.

We have seen that all the non-relativistic algebras and their bilinear forms 
can be obtained from the relativistic one by contractions. 
In addition, it is also sufficient to consider the chiral sector $sl(2)\oplus u(1)$.
At the level of Lie algebra, there is no relation between $sl(2)$ and $u(1)$,
hence their bilinear forms  \eqref{rel bil} are also independent. 

On the other hand, one simple way to introduce an invariant bilinear form is through 
an invariant linear form, i.e. trace:
\be
	\la X, Y\ra =\tr(X\, Y).
\ee
A well-defined trace requires an associative product between  $X$ and $Y$.
This can be done by representing $X$ and $Y$ as matrices. But the price to pay is that
we have to work with a larger space because the associative products of the original elements will generate 
additional elements. The enlargement can be minimized by choosing the smallest representation.
For $sl(2)$, the smallest representation is the fundamental representation, 
which is two dimensional and requires only one more element, i.e. the identity, to have the associative structure.
The associative product of the relativistic chiral algebra reads
\be
	\cL_a\,\cL_b=\eta_{ab}\,\cI+\e_{ab}{}^c\,\cL_c,
	\qquad
	\tilde \cL_a\,\tilde \cL_b=\eta_{ab}\,\tilde \cI+\e_{ab}{}^c\,\tilde \cL_c,
	\label{associative prod}
\ee
and 
we recover the bilinear form \eqref{LL-tr} using
\be
	\tr(\cI)=\l,\quad
	\tr(\tilde \cI)=\tilde \l,
	\quad \tr(\cL_a)=\tr(\tilde\cL_a)=0.
	\label{h trace}
\ee
There is no a priori reason to identify the identities $\cI$ and $\tilde \cI$ with the $u(1)$ generators $\cZ$ and $\tilde \cZ$. 
However, if they coincide up to some factors $z(c)$ and $\tilde z(c)$, the coefficients $\l, \tilde \l$ and $\z, \tilde\z$ in \eqref{rel bil}, \eqref{LL-tr} will be related as
\be
	\cZ=z(c)\,\cI \quad \Rightarrow \quad  \z=z^2(c)\,\l,
	\label{Z-id}
\ee
\be
	\tilde\cZ=\tilde z(c)\,\tilde\cI \quad \Rightarrow \quad \tilde \z=\tilde z^2(c)\,\tilde\l. 
	\label{Z-id-tilde}
\ee
These relations together with \eqref{l z} fix the form of the functions $z(c)$ and $\tilde z(c)$ as
\be
	z^2(c)=\frac{c^2\ell+\b/2}{c^2\ell-\b/2},\qquad
	\tilde z^2(c)=\frac{c^2\tilde \ell+\tilde \b/2}{c^2\tilde \ell-\tilde \b/2}\,.
	\label{z-k-l}
\ee
Therefore, we fully recover the non-relativistic algebra $\mathfrak h$ and $\tilde{\mathfrak h}$, and 
their bilinear forms using the associative structure of the relativistic counterpart, $gl(2)$.
As we mentioned in the introduction,
the associative structure of the isometry algebras plays a central role in the color decoration of 
three-dimensional gravity.
From this observation, we are driven to look more closely the possible interplay 
between non-relativistic and color deformations of gravity
in three dimensions.

\section{Colored Gravity and its Non-relativistic Limits}\label{sec-non-relativistic-colored-alg}

If a theory contains more than one graviton and they are related by internal symmetry, 
the global symmetry should be extended 
from the usual isometry $\mathfrak{g}_i$ to a
color-decorated one $\mathfrak{g}_i \otimes \mathfrak g_c$,
where $\mathfrak g_c$ is a finite vector space generated by the color label.
This leads to severe algebraic constraints: 
there is only trivial possibility if we assume the color-decorated algebra
does not include additional generators, 
which means extra fields are necessary \cite{Gwak:2015vfb,Gwak:2015jdo}.
Even if we find a good candidate of colored gravity algebras by including a reasonable amount of additional fields,
the construction of equations of motion or/and actions is not guaranteed to be straightforward. 
In three dimensions, the situation is much simpler.
First, thanks to the Chern-Simons formulation, we obtain the action directly from a given algebra.
Second, a color-decorated algebra can be constructed by 
extending the isometry algebra with only two additional $u(1)$ generators.

Technically speaking, we want to find a Lie algebra structure on 
the tensor-product vector space $\mathfrak{g}_i \otimes \mathfrak g_c$\,. 
The Lie brackets between two elements of the above space may be written as
\be
	[\,M_{X}\otimes \bm \cT^{I},M_{Y}\otimes \bm \cT^{J}\,]
	=\frac12\,[\,M_{X},M_{Y}\,]\otimes \{\,\bm \cT^{I} ,\bm \cT^{J}\,\}
	+\frac12\,\{\,M_{X}, M_{Y}\,\} \otimes [\,\bm \cT^{I},\bm \cT^{J}\, ] \,,
	\label{color bracket}
\ee
where $M_X$ and $M_Y$ belong to $\mathfrak{g}_i$ whereas $\bm \cT^I$ and $\bm \cT^J$ to $\mathfrak{g}_c$\,.
Note that we have used both the commutator and anti-commutator of the generators of each space. 
Therefore, in order to have well-defined Lie brackets \eqref{color bracket}, 
we should begin with two associative algebras $\mathfrak{g}_i$ and $\mathfrak{g}_c$\,.
For the color algebra $\mathfrak{g}_{c}$\,, we use the matrix algebra $u(N)$.\footnote{In principle, any finite-dimensional associative algebra can 
be used as the color algebra.} 
For the isometry algebra $\mathfrak{g}_{i}$\,,
the minimal choice is   $gl(2)\oplus gl(2)$.
This is precisely the same as the setting \eqref{assoc} for the doubly extended 
NH algebra.

\subsection{Chiral sector}

Let us start from the non-relativistic limits of one chiral sector of the colored gravity algebra.
It is generated by 
the following generators,
\be
	\cZ,\qquad \cL_a\,,\qquad \bm \cT^I,\qquad \bm\cL_a^I,
\ee
where $\cL_a$ and $\cZ$ form the original relativistic algebra $sl(2)\oplus u(1)=gl(2)$
and  $\bm \cT^I$ are the $su(N)$ generators:
\be
	[\bm \cT^I,\bm \cT^J]=2\,i\,f^{IJ}{}_K\,\bm \cT^K.
\ee
The generators $\bm\cL_a^I$ are the color decoration of the spin-two generators and 
they transform covariantly under $sl(2)$ and $su(N)$:
\be
	[\cL_a,\bm\cL^I_b]=2\,\e_{ab}{}^c\,\bm\cL^I_c,
	\qquad
	[\bm \cT^I,\bm\cL^J_a]=2\,i\,f^{IJ}{}_K\,\bm\cL^K_a,
\ee
but the Lie brackets between them are non-trivial:
\be\label{Lie-bra-colored-isometry}
	[\bm\cL_a^I,\bm\cL^J_b]=
	2\,i\,\eta_{ab}\,f^{IJ}{}_K\,\bm \cT^K+
	2\,\e_{ab}{}^c\left(\frac 1 N\d^{IJ}\,\cL_c+g^{IJ}{}_K\,\bm\cL^K_c\right).
\ee
In \eqref{Lie-bra-colored-isometry}, we used the associative algebra of $u(N)$
\be
\bm \cT^I\,\bm \cT^J=\frac 1 N\,\d^{IJ}I
+g^{IJ}{}_K\,\bm \cT^K
+i\,f^{IJ}{}_K\,\bm \cT^K,
\ee
where $g_{IJK}$ is totally symmetric and $f_{IJK}$ totally antisymmetric.  
We assign its bilinear form as
\be
	\la \bm\cT^I,\bm\cT^J\ra
	=\tr(\bm\cT^I\,\bm\cT^J)=\delta^{IJ},
	\label{TT bf}
\ee
where we used $\tr(I)=N$ while $\tr(\bm \cT^I)=0$.

We want to construct the non-relativistic analogs of the above colored gravity algebra,
which we assume to have the same number of generators as the relativistic one. 
The non-relativistic algebra is spanned by
\be
	Z, \quad L_i, \quad K,\quad \bm \cT^I, \quad \bm L_i^I,\quad \bm K^I.
\ee
Their Lie brackets depend on the non-relativistic limits.

\subsubsection*{Color decoration after contraction}

One of the possible ways to obtain the non-relativistic colored gravity algebra
is by colored-decorating the doubly extended NH algebra.
For that, the latter algebra should admit an associative algebra structure.
In the previous section, we have shown that it can be obtained 
from an associative relativistic algebra by a suitable contraction.
However, we have not examined the
full consistency of the associative structure of the resulting non-relativistic algebra.
An associative product can be split into commutators, which reproduce the Lie brackets,
and anti-commutators.
Therefore, we have to check whether the contraction limit \eqref{chiral def} is equally well-defined for the anti-commutators.
From \eqref{associative prod}, \eqref{Z-id} and \eqref{z-k-l}, 
most of the anti-commutators are regular in the contraction limit: 
\be
	\{K,\,L_i\}=2\,L_i,
	\quad
	\{ K, \,Z\}=2\,Z,
	\quad
	\{L_i,\,L_j\}=2\,\delta_{ij}\,Z,
	\quad
	\{L_i, \,Z\}=0,
	\quad
	\{Z,\,Z\}=0.
	\quad 
\ee
However, before taking the contraction limit, 
the $\{K,K\}$ anti-commutator reads
\be\label{anti-com-KK}
	\{ K, K \}=2\,(K-2\,c^2\,Z).
\ee
In the $c\to \infty$ limit, the anti-commutator \eqref{anti-com-KK} becomes ill-defined, 
so the corresponding associative product is problematic.
The contraction of the $gl(2)$ gives a good Lie algebra structure and bilinear form or trace,
but not the fully associative algebra structure. 

This issue does not affect the Chern-Simons action of the doubly extended NH gravity. 
The reason we consider the anti-commutator $\{K,K\}$ here 
is that from \eqref{color bracket} they appear in the Lie bracket of its color-decorated counterparts, 
$[\bm K^I,\bm K^J]$. 
We can avoid the divergence by defining the colored generators as
\be
	\bm K^I=\frac1c\,K\otimes \bm \cT^I,
	\qquad
	\bm Z^I=Z\otimes \bm \cT^I,
\ee
which lead to a regular commutator
\be
	[\bm K^I,\bm K^J]
	=-4\,i\,f^{IJ}{}_K\,\bm Z^K.
\ee
Concerning the color decoration of $L_i$ generators, 
we consider two options:
\begin{itemize}
\item
In the first option, by defining them as
\be
	\bm L^I_i=\frac1{c}\,L_i\otimes \bm \cT^I,
\ee
we have
\be
	[\bm K^I,\,L_i]=[K,\,\bm L_i^I]=2\,\e_{ij}\,\bm L_j^I.
\ee
\item
In the second option, with the definition
\be
	\bm L^I_i=L_i\otimes \bm \cT^I,
\ee
we obtain different Lie brackets,
\be
	[K,\,\bm L_i^I]=2\,\e_{ij}\,\bm L_j^I,
	\qquad
	[L_i, \bm L^I_j]=2\,\e_{ij}\,\bm Z^I,
\ee
\be
	[\bm L^I_i, \bm L^J_j]=
	2\,\epsilon_{ij}\left(\frac 1 N\,\e_{ij}\,\delta^{IJ}\,Z
	+2\,g^{IJ}{}_K \bm Z^K\right)
	+2\,i\,\delta_{ij}\,f^{IJ}{}_K \,\bm Z^K.
\ee
\end{itemize}
The rest of the Lie brackets vanish in both cases.

In the above paragraph, we obtained two non-relativistic colored gravity algebras with suitable contractions. 
However, it turns out that they do not have non-degenerate invariant bilinear form
due to the additional contraction related to the color decoration.
This problem can be fixed by considering the mixing between the two chiral algebras, 
which will be explained in section \ref{non-rel-alg}.
In the next subsection, 
we will introduce the generalized \.In\"on\"u-Wigner contraction,
whose concept is crucial to the systematic construction of non-relativistic algebras from the relativistic ones.

\subsection{Generalized \.In\"on\"u-Wigner contraction}

\.In\"on\"u-Wigner contraction \cite{Inonu:1953sp}  is a procedure 
for generating Lie algebras from a starting Lie algebra 
by taking certain limits of the group-contraction parameters. 
The contraction procedure involves singular changes of bases, 
so the contracted algebra is not isomorphic to the original algebra. 
If a Lie algebra $\mathfrak g$ has a subalgebra $\mathfrak h$, we can decompose $\mathfrak g$ into
\be
\mathfrak g=\mathfrak h\oplus\mathfrak i.
\ee
We then rescale the generators of $\mathfrak i$ by a contraction parameter $c$ 
and introduce the new generators as
\be
\mathfrak j= \frac 1 c\,\mathfrak i.
\ee
The Lie brackets become
\be
[\mathfrak h,\mathfrak h]\subset \mathfrak h,\qquad
[\mathfrak h,\mathfrak j]\subset \frac 1 c \, \mathfrak h \oplus \mathfrak j,\qquad
[\mathfrak  j,\mathfrak j]\subset \frac 1 {c^2} \,\mathfrak h \oplus \frac 1 c\, \mathfrak j. 
\ee
In the singular limit where $c\to\infty$,
we obtain a new Lie algebra,
\be
	\cG(\mathfrak g)=\mathfrak h\oplus \mathfrak j,
\ee
with the Lie brackets,
\be
[\mathfrak h,\mathfrak h]\subset \mathfrak h,\qquad
[\mathfrak h,\mathfrak j]\subset \mathfrak j,\qquad
[\mathfrak j,\mathfrak j]\subset \emptyset , 
\ee
where $\mathfrak j$ corresponds to an Abelian ideal. 
As an example, the Poincar\'e algebra can be obtained from the (A)dS algebra 
by sending the radius of curvature to infinity, 
and the translation generators in the Poincar\'e algebra generate an Abelian ideal.

\subsubsection*{Three-level  contraction}

The standard contraction procedure can be generalized. 
Assume the original Lie algebra $\mathfrak g$ admits a decomposition into three parts:
\be
\mathfrak g=\mathfrak i_0\oplus \mathfrak i_1 \oplus \mathfrak i_2,
\ee
with the condition,
\be
	[\mathfrak i_0,\mathfrak i_0]\subset \mathfrak i_0,
	\qquad
	[\mathfrak i_0,\mathfrak i_1]\subset \mathfrak i_0 \oplus \mathfrak i_1.
\ee
If we rescale the generators $\mathfrak i_1$ and $\mathfrak i_2$ by $c^{-1}$ and $c^{-2}$, respectively:
\be
\mathfrak j_0=\mathfrak i_0,\qquad
\mathfrak j_1=\frac 1 c\,\mathfrak i_1,\qquad
\mathfrak j_2=\frac 1 {c^2}\,\mathfrak i_2,
\ee
the contraction limit $c\to\infty$ gives a new algebra, 
\be
	\cG(\mathfrak g)=\mathfrak j_0\oplus \mathfrak j_1\oplus \mathfrak j_2,
\ee
with the Lie brackets,
\be
[\mathfrak j_0,\mathfrak j_0]\subset \mathfrak j_0,\quad
[\mathfrak j_0,\mathfrak j_1]\subset \mathfrak j_1,\quad
[\mathfrak j_0,\mathfrak j_2]\subset \mathfrak j_2,\quad
[\mathfrak j_1,\mathfrak j_1]\subset \mathfrak j_2,
\ee
and
\be
[\mathfrak j_1,\mathfrak j_2] \subset \emptyset,\qquad
[\mathfrak j_2,\mathfrak j_2] \subset \emptyset.
\ee
Note that $\mathfrak j_2$ and $\mathfrak j_1+\mathfrak j_2$ are two different ideals of the contracted Lie algebra. 
This contraction
actually covers all the non-relativistic limits considered in this paper.

\subsubsection*{Multi-level contraction}

We can further generalize the contraction procedure by considering 
arbitrary many ``levels'', 
i.e. several integer rescaling dimensions. 
The starting point is again a decomposition of the original algebra
\be
	\mathfrak g=\mathfrak i_0\oplus \mathfrak i_1 \oplus \cdots \oplus \mathfrak i_{L},
	\label{multi decomp}
\ee
with the condition
\be
	[\mathfrak i_m,\mathfrak i_n]\subset \mathfrak i_0\oplus \cdots \oplus \mathfrak i_{m+n},
	\qquad
	\mathfrak i_{n>L}=\emptyset.
	\label{i cond}
\ee
After the rescaling
\be
	\mathfrak j_k=\frac 1 {c^k}\,\mathfrak i_k,
\ee 
and the contraction $c\rightarrow \infty$, we obtain a new algebra,
\be
	\cG(\mathfrak g)
	=\mathfrak j_0 \oplus \mathfrak j_1
	\oplus \cdots \oplus \mathfrak j_L,
\ee
with the Lie brackets
\be
	[\mathfrak j_m, \mathfrak j_n]\subset \mathfrak j_{m+n},
	\qquad \mathfrak j_{n>L}=\emptyset. 
	\label{grading}
\ee 
When the levels of the elements are higher than $L/2$, 
their commutators will vanish 
as the maximum is $L$. 
At least half of the ideals $\mathfrak j_n\oplus \mathfrak j_{n+1}\oplus \cdots
\oplus \mathfrak j_L$ are Abelian.
We need $L>1$ to have a non-abelian ideal. 

Mathematically, the condition \eqref{i cond} is nothing but a filtration structure,
for which it is more nature to work with
\be
	\mathfrak f_n=\mathfrak i_0\oplus \cdots \oplus \mathfrak i_n.
\ee
Then, the decomposition \eqref{multi decomp} and the condition \eqref{i cond} can be recast as
the definition of the filtered Lie algebra:
\be
	\mathfrak f_0 \subset \cdots\subset \mathfrak f_{L-1}  \subset \mathfrak f_L=\mathfrak g,
	\quad 
	{\rm with}\quad 
	[\mathfrak f_m, \mathfrak f_n]\subset \mathfrak f_{m+n}.
\ee
For a given filtered (Lie) algebra $\mathfrak g$, 
there exists a graded (Lie) algebra $\cG(\mathfrak g)$
with \eqref{grading}.
The latter is referred to as \emph{associated graded (Lie) algebra}.
In the mathematical physics literature, this generalized \.In\"on\"u-Wigner contraction has been studied in \cite{Weimar-Woods}.

An example of multi-level contractions can be found in higher spin algebras. 
By definition, the vector space of a higher spin algebra consists of
the spaces $\mathfrak{K}_{s}$ of spin $s$ Killing tensors,
\be
	\mathfrak{g}_{HS}
	=\bigoplus_{s}\mathfrak{K}_{s}.
\ee
Each $\mathfrak{K}_{s}$ is generated by the traceless tensor $M_{a_1\cdots a_{s-1},b_1\cdots b_{s-1}}$
with the permutation symmetry of two-row Young diagram $(s-1,s-1)$.
The Lie brackets of the higher spin algebra satisfy
\be\label{higher-spin-bra}
	[\mathfrak{K}_{s_1}, \mathfrak{K}_{s_2}]
	\subset \mathfrak{K}_{s_1+s_2-2}\oplus \mathfrak{K}_{s_1+s_2-3} \oplus \cdots.
\ee
The spins on the right hand side of \eqref{higher-spin-bra} is bounded from above
because one cannot construct $(r,r)$ tensors with $r\ge s_1+s_2-2$
in terms of $M_{(s_1-1,s_2-1)}$ and $M_{(s_2-1,s_2-1)}$.\footnote{The case of $r=s_1+s_2-2$ is ruled out 
by the antisymmetricity of Lie bracket.}
Hence, taking $\mathfrak{i}_{s-2}=\mathfrak{K}_s$, one can 
see that the higher spin algebra enjoys the filtration structure \eqref{i cond}.
An unusual point is that a higher spin algebra is typically infinite dimensional, 
so it involves infinitely many levels ($L=\infty$).
In three dimensions, one can consider finite dimensional higher spin algebras.
For instance, $sl(N,\mathbb{R})$ is isomorphic to the chiral part
of the higher spin algebras involving spin $2$ to $N$. 
The contracted higher spin algebra still satisfies all the higher spin covariance conditions,
but does not admit an invariant bilinear form.
It would be interesting to check whether a central extension of this contracted higher spin algebra
may admit a non-degenerate bilinear form, 
as in the case of three-dimensional non-relativistic spin-two algebras.

\subsection{Non-relativistic colored gravity algebra}\label{non-rel-alg}

In this subsection, we will study the non-relativistic limits of the three-dimensional colored gravity algebra
using the generalized \.In\"on\"u-Wigner contraction.
The guiding principle
is that the limiting algebra should have a non-degenerate bilinear form. 
This non-degeneracy is related to the physical requirement 
that the kinetic terms are present.\footnote{In 
a more general setting, we may consider degenerate bilinear form. 
Then some fields will not have kinetic terms and only appear in the cubic terms.}
As we have seen before, 
this condition cannot be met in the chiral sector,
so we will work with the full algebra from now on.

We shall define the generators of the non-relativistic colored gravity algebras 
in terms of the relativistic ones and the contraction parameter $c$.
For the color-singlet generators, 
we will use the same contraction ansatz as the doubly extended NH algebra 
$\mathfrak g^{++}_{\L}$, i.e. \eqref{PG}, \eqref{HJ mix} and \eqref{MS mix}. 
This choice corresponds to the three-level  ansatz:
\ba
	&& \mathfrak{j}_0\quad : \quad 
	\cZ+\cL_0\,,\quad \tilde \cZ+\tilde \cL_0\,,\nn
	&& \mathfrak{j}_1\quad : \quad
	\frac{\cL_i}c,\quad \frac{\tilde\cL_i}c\,, \nn
	&& \mathfrak{j}_2\quad : \quad
	\frac{\cZ-\cL_0}{c^2}\,, \quad \frac{\tilde\cZ-\tilde\cL_0}{c^2}\,.
\ea
The non-singlet generators of the colored gravity algebra
should be appended to $\mathfrak{j}_0$, $\mathfrak{j}_1$ and $\mathfrak{j}_2$
such that the contracted algebra admits a non-degenerate bilinear form.
From the symmetry consideration, the bilinear form of the non-singlet sector generators
are determined by that of the singlet sector as
\be
	\la a\otimes \bm\cT^I, b\otimes \bm\cT^J\ra
	=\la a,b\ra \,\delta^{IJ}\,,
\ee
where $a$ and $b$ are elements of $gl(2)\oplus gl(2)$ and their bilinear form is given in Section \ref{sec: inv bf}\,.
Let us summarize it here for the reader's convenience: 
due to the associative structure, 
the bilinear form should be diagonal, i.e. $\la \cZ, \tilde \cZ\ra=\g=0$, then we have
\ba
	&& \la \cZ,\cZ\ra=c^2\,\ell+\b/2\,,
	\qquad \la \cL_a,\cL_b\ra=\eta_{ab}\,(c^2\,\ell-\b/2)\,,\nn
	&& \la \tilde\cZ,\tilde\cZ\ra=c^2\,\tilde\ell+\tilde \b/2\,,
	\qquad \la \tilde \cL_a,\tilde \cL_b\ra=\eta_{ab}\,(c^2\,\tilde\ell-\tilde \b/2)\,.
\ea
As one can see, they all scale as $c^2$ hence diverge unless we
rescale them appropriately.
About the spatial part of non-singlet generators, we take the analogous rescaling as the singlet ones:
\be 
\mathfrak{j}_1\quad : \quad
 \frac{\cL_i\otimes \bm \cT^I}{c},\quad \frac{\tilde \cL_i\otimes \bm \cT^I}{c},
\ee
hence these generators belong to $\mathfrak{j}_1$\,.
About the remaining generators, there are a few consistent possibilities.
In the classification, an important guideline is the fact that  $\mathfrak{j}_0$ must form a subalgebra
for a consistent contraction. 
Moreover, due to the presence of $su(N)$ part, the $gl(2)\oplus gl(2)$ part of the non-singlet elements in $\mathfrak{j}_0$
should be closed under anti-commutator:
let us write the non-singlet part of $\mathfrak{j}_0$ as $\mathfrak{A}\otimes \bm \cT^I$, 
then $\mathfrak{A}$ should satisfy
\be
	\{\mathfrak{A},\mathfrak{A}\}\subset \mathfrak{A}\,,
\ee
where $\mathfrak{A}$ is a subspace of $gl(2)\oplus gl(2)$. 
With the above condition, let us examine what are the possible candidates 
for the $\mathfrak{A}$ elements.
The first guess is the combinations analogous to the singlet sector:
\be
 (\cZ+\cL_0)\otimes \bm \cT^I,\qquad (\tilde \cZ+\tilde \cL_0)\otimes \bm \cT^I,
\ee
but one can easily see that neither $\cZ+\cL_0$ nor $\tilde \cZ+\tilde \cL_0$ 
is closed under anti-commutator.
In addition, we cannot consider $\cZ$ or $\tilde\cZ$ because their bilinear form will diverge.
In fact, the only possible way to have finite bilinear form in $\mathfrak{j}_0$ part --- that is, without rescaling
the generators --- is to cancel the $c^2$ terms: for instance, $\la \cZ+\cL_0,\cZ+\cL_0\ra
=\la \cZ,\cZ\ra+\la\cL_0,\cL_0\ra=(c^2\,\ell+\b/2)-(c^2\,\ell-\b/2)=\b$  in the singlet sector.
Noticing this, we can see 
\be
	\mathfrak{j}_0\quad :\quad (\cZ+\tilde \cZ)\otimes \bm \cT^I
	\label{ZZ0}
\ee
has a finite bilinear form: $\la \cZ+\tilde \cZ,\cZ+\tilde \cZ\ra=\frac{\b+\tilde \b}2$, and it is closed 
under anti-commutator: $(\cZ+\tilde \cZ)^2=\cZ+\tilde \cZ$.
The similar combination $\cZ-\tilde \cZ$ has a finite bilinear form but is not closed.
For the non-degeneracy of the bilinear form, the choice \eqref{ZZ0} should be 
accompanied by
\be
	\mathfrak{j}_2\quad :\quad \frac{(\cZ-\tilde \cZ)\otimes \bm \cT^I}{c^2}.
	\label{ZZ2}
\ee
Now there remain only two generators $\cL_0$ and $\tilde \cL_0$\,,
and there are two non-trivial choices:
\be
	\mathfrak{j}_0\ :\  (\cL_0\pm \tilde \cL_0)\otimes \bm \cT^I,
	\qquad
	\mathfrak{j}_2\ :\  \frac{(\cL_0\mp \tilde \cL_0)\otimes \bm \cT^I}{c^2}.
	\label{L+-}
\ee
Note that both signs give consistent contractions.
There also exist rather trivial options:
{\it i)} the entire non-singlet generators are in $\mathfrak{i}_1$; 
{\it ii)} 
we have \eqref{ZZ0} and \eqref{ZZ2} but $\cL_0\otimes \bm\cT^I$ and $\tilde \cL_0\otimes \bm\cT^I$ are in $\mathfrak{i}_1$.
The latter two cases are mathematically consistent options, but
they treat the color-decorated isometry generators $\cL_a\otimes \bm\cT^I$ 
in the Lorentz invariant manner, hence they are unnatural for 
the interpretation as non-relativistic physics.
In the following, we consider only the ``non-trivial'' cases. See
Appendix \ref{sec: case01} for the other two cases.
Another implicit assumption is that the $su(N)$ symmetry is not broken in the non-relativistic limits. 
There will be more possibilities if we relax this assumption.

\section{Chern-Simons Formulation }\label{sec-non-relativistic-colored-action}

In this section, we derive the actions of 
the non-relativistic colored gravity theories. 
There are two ways to compute these non-relativistic actions. 
The first way is to directly evaluate the Chern-Simons actions 
whose gauge fields take values in the non-relativistic gauge algebras. 
The second way is to take the non-relativistic limits of the relativistic Chern-Simons action 
according to the contractions. 
We will use the first method in the current section and 
discuss the second one in Appendix \ref{contraction-action}.

\subsection{Gauge algebras}

Before starting our discussion, let us summarize
the algebraic structure of non-relativistic colored gravity symmetry.
For a more clear physical interpretation, we use ``$\cJ$, $\cP$'' bases
rather than the chiral and anti-chiral bases.
Since the singlet sector is identical to the doubly extended NH algebra, 
we directly start with the non-singlet ones.
The contractions we have considered in the previous section are
\be
\bm C^I=\cZ_{\cJ}\otimes \bm \cT^I,\quad
\bm D^I=\frac 1 {c^2}\,\cZ_{\cP}\otimes \bm \cT^I,
\ee
\be
\bm G_i^I=\frac 1 {c} \cJ_i\otimes \bm \cT^I,\quad 
\bm P_i^I=\frac 1 {c} \cP_i\otimes \bm \cT^I.
\ee
These generators transform covariantly under the NH symmetry as
\be
[J,\,\bm G_i^I]
=2\,\epsilon_{ij}\,\bm G_j^I,
\,\quad
[J,\,\bm P_i^I]
=2\,\epsilon_{ij}\,\bm P_j^I,
\ee
\be
[H,\,\bm G_i^I]
=2\,\epsilon_{ij}\, \bm P_j^I,
\,\quad
[H,\,\bm P_i^I]
=-2\,\L\,\epsilon_{ij}\,\bm G_j^I,
\ee
and they have non-trivial Lie brackets among themselves:
\be
[\bm C^I,\, \bm C^J]=2\,i\, f^{IJ}{}_K\, \bm  C^K,
\quad
[\bm C^I,\,\bm D^J]=2\,i\, f^{IJ}{}_K\, \bm D^K,
\ee
\be
[\bm C^I,\,\bm G_i^J]=2\,i\, f^{IJ}{}_K\, \bm G_i^K,
\quad
[\bm C^I,\, \bm P_i^J]=2\,i\, f^{IJ}{}_K\,  \bm P_i^K.
\ee
Besides $\bm C^I, \bm D^I, \bm G_i^I$ and $\bm P_i^I$,
we also have the temporal and central generators.
The two possibilities of the sign in \eqref{L+-}
are translated into the following two cases:
\be
 {\rm Type\ A}\quad :\quad \bm J^I=\cJ_0\otimes \bm \cT^I,\quad 
\bm M^I=\frac 1 {c^2}\, \cP_0\otimes \bm \cT^I,
\ee
\be
 {\rm Type\ B}\quad :\quad 
\bm H^I=\cP_0\otimes \bm \cT^I,\quad
\bm S^I=\frac 1 {c^2}\,\cJ_0\otimes \bm \cT^I. 
\ee
Only one of the two central generators in the doubly extended NH algebra admits color decoration. 
Let us present the Lie brackets of the two cases.

\subsubsection*{Type A}
The non-zero commutators involving one singlet and one non-singlet are
\be
[\bm J^I,\,G_i]
=2\,\epsilon_{ij}\, \bm G_j^I
,\quad
[\bm J^I,\,P_i]
=2\,\epsilon_{ij}\, \bm P_j^I,
\ee
\be
[ \bm G_i^I,\, P_j]
=[ G_i,\,\bm P_j^I]
=-2\,\epsilon_{ij}{}\, \bm  M^I.
\ee
The non-zero commutators involving only the non-singlet sectors are
\be
[\bm C^I,\,\bm J^J]=2\,i\, f^{IJ}{}_K\, \bm J^K,
\quad
[\bm C^I,\, \bm M^J]=[\bm D^I,\,\bm J^J]
=2\,i\, f^{IJ}{}_K\,  \bm M^K,
\ee
\be
[\bm J^I,\,\bm J^J]
=
-2\,i\, f^{IJ}{}_K\, \bm C^K,
\quad
[\bm J^I,\, \bm M^J]
=
-2\,i\,f^{IJ}{}_K\,\bm D^K,
\ee
\be
[\bm J^I,\,\bm G_i^J]
=
2\,\epsilon_{i j}\left(\frac 1 N\d^{IJ} G_j+g^{IJ}{}_K\,\bm G_j^K\right)
,
\ee
\be
[\bm J^I,\, \bm P_i^J]
=
2\,\epsilon_{i j}\left(\frac 1 N\d^{IJ} P_j+g^{IJ}{}_K\,\bm P_j^K\right),
\ee
\be
[\bm G_i^I,\,
\bm G_j^J]=
-\frac 1 \L \,[\bm P_i^I,\,
\bm P_j^J]
=
\frac 2{N}\,\epsilon_{ij}\,\d^{IJ}S,
\ee
\be
[\bm G_i^I,\, \bm P_j^J]
=
\frac 2 {N}\,\epsilon_{ij}\,\d^{IJ}\,M
-2\,\epsilon_{ij}\,g^{IJ}{}_K\, \bm M^K
+
2\,i\,\d_{ij}\,f^{IJ}{}_K\,\bm D^K.
\ee

\subsubsection*{Type B}

The non-zero commutators containing one singlet and one non-singlet sectors are
\be
[\bm H^I,\,G_i]
=2\,\epsilon_{ij}\, \bm P_j^I
,\quad
[\bm H^I,\, P_i]
=-2\,\L\,\epsilon_{ij}\,\bm G_j^I,
\ee
\be
[G_i,\,  \bm G_j^I]
=-\frac 1 \L\, [ P_i,\,\bm P_j^I]
=-2\,\epsilon_{ij}{}\, \bm  S^I.
\ee
The non-zero Lie brackets involving only the non-singlet sector are
\be
[\bm C^I,\, \bm H^J]=2\,i\, f^{IJ}{}_K\,  \bm H^K,
\quad
[\bm C^I,\,\bm S^J]=-\frac 1 \L [\bm D^I,\,
 \bm H^J]=
 2\,i\, f^{IJ}{}_K\, \bm S^K,
\ee
\be
[ \bm H^I,\,
 \bm H^J]
=
2\,i\,\L\, f^{IJ}{}_K\,\bm C^K
,
\quad
[\bm S^I,\, \bm H^J]
=
-2\,i\,f^{IJ}{}_K\,\bm D^K,
\ee
\be
[\bm H^I,\,\bm G_i^J]
=
2\,\epsilon_{i j}\left(\frac 1 N\d^{IJ} P_j+g^{IJ}{}_K\,\bm P_j^K\right)
,
\ee
\be
[\bm H^I,\, \bm P_i^J]
=
-2\,\L\,\epsilon_{i j}\left(\frac 1 N\,\d^{IJ} G_j+g^{IJ}{}_K\,\bm G_j^K\right),
\ee

\be
[\bm G_i^I,\,
\bm G_j^J]=
-\frac 1 \L\, [\bm P_i^I,\,
\bm P_j^J]
=
\frac 2 {N}\,\epsilon_{ij}\,\d^{IJ}S
-2\,\epsilon_{ij}\,g^{IJ}{}_K \bm S^J,
\ee
\be
[\bm G_i^I,\, \bm P_j^J]
=
\frac 2 {N}\,\epsilon_{ij}\,\d^{IJ}M
+
2\,i\,\d_{ij}\,f^{IJ}{}_K\,\bm D^K.
\ee

\subsubsection*{Bilinear form}

Let us also remind the reader that the invariant bilinear form reads
\be
	\la J, M\ra=\la S, H\ra=1\,,\qquad
	\la P_i,\,G_j\ra=\delta_{ij}\,,
\ee
\be
	\la \bm C^I ,\,\bm D^J \ra=\d^{IJ},\qquad
	\la \bm  P_i^I,\, \bm G_j^J \ra=\d_{ij}\,\d^{IJ},
\ee
and
\be
	\la \bm  J^I ,\,\bm M^J \ra=-\d^{IJ},\qquad\
		\la \bm  H^I ,\,\bm S^J \ra=-\d^{IJ},
\ee
for type A and B, respectively.
Here, we took the choice $\b=\tilde\b=\g=0$ for simplicity.

\subsection{Actions}\label{explicit-action}

Now we consider the Chern-Simons actions of the non-relativistic colored gravities. 
The one-form gauge fields are
 \ba\label{one-form-A}
\text{Type A}\quad : \quad
 \cA&=&j J+s\, S+h\, H+m\, M+\O^i\, G_i+E^i\, P_i
 \nn&&
+\s_I\,\bm C^I+ \varsigma_I\, \bm D^I
+ \vartheta_I\, \bm J^I+ \varpi_I\, \bm M^I 
+ \theta_I^i\, \bm G^I_i+ \pi_I^i\,\bm P^I_i,
\ea
 \ba\label{one-form-B}
 \text{Type B}\quad : \quad
 \cA&=&j J+s\, S+h\, H+m\, M+\O^i\, G_i+E^i\, P_i
 \nn&&
+\s_I\,\bm C^I+ \varsigma_I\, \bm D^I
+ \vartheta_I\, \bm S^I+ \varpi_I\, \bm H^I 
+ \theta_I^i\, \bm G^I_i+ \pi_I^i\,\bm P^I_i,
\ea
which correspond to the non-relativistic colored gravity algebras constructed in Section \ref{non-rel-alg}.
It is now sufficient to plug the above into the Chern-Simons action,
\be
	S=\frac{\k}{4\pi}\,\int \la \cA \wedge d\cA\ra
	+\frac23\,\la \cA\wedge \cA\wedge \cA\ra.
\ee 
In terms of the component fields, this action can be decomposed into several pieces:
\be
	S=S_{\text{gravity}}+S_{\text{gauge}}+S_{\text{matter}}.
	\label{Action 3}
\ee
Here, $S_{\text{gravity}}$ is the usual uncolored non-relativistic action,
\ba\label{gra-action}
S_{\text{gravity}}&=&
\frac {\kappa\,N} {2\pi}
\int E^i\wedge \nabla\,\O_i+m\wedge d\,j
+h\wedge [d\,s+\epsilon_{ij}\,(\O^i\wedge \O^j
-\L\,E^i\wedge E^j)],
\ea
where the covariant derivative $\nabla$ acts on ``spatial'' one-form $\o_i$ as
\be
	\nabla\,\o_i=d\,\o_i+\epsilon_{ij}\left(\o^j\wedge j-j\wedge \o^j\right).
\ee
About the other pieces of the action,
it is convenient to introduce the $su(N)$-valued fields as 
\be
\bm\s=\s_I\,\bm \cT^I,\quad \bm \varsigma=\varsigma_I\, \bm \cT^I,\quad
\bm\vartheta=\vartheta_I\, \bm \cT^I,\quad \bm \varpi=\varpi_I\, \bm \cT^I,\quad
\bm \theta^i=\theta_I^i\, \bm \cT^I,\quad \bm\pi^i=\pi_I^i\,\bm \cT^I,
\ee
and write the action in terms of $\text{Tr}$, the trace in the adjoint representation.
Then, the gauge field part $S_\text{gauge}$ takes the simple form,
\be
S_{\text{gauge}}
=
\frac {\kappa } {2\pi}
\int
 \text{\tr }\left[
\bm  \varsigma\wedge (d\bm \s+\bm\s\wedge\bm\s)
\right].
\label{S gauge}
\ee
The matter field part $S_\text{matter}$ 
depends whether we consider the type A or B:
\be
	S_\text{matter}=S_\text{matter,0}+S_\text{type A/B}\,.
\ee
The common part $S_\text{matter,0}$ is
\be
S_{\text{matter,0}}=
\frac {\kappa } {2\pi}
\int
 \tr\left[
 -\bm \vartheta\wedge D\,\bm  \varpi
+ \bm \theta^i\wedge D\,\bm \pi_i
+ \e_{ij}\,h\wedge(\bm \theta^i\wedge\bm \theta^j-\L\, \bm\pi^i \wedge \bm\pi^j)\right],
\label{S matter 0}
\ee 
where the covariant derivative $D$ acts on 
the $su(N)$ valued  one-form $\bm\o$ as
\be
D\,\bm \o=\nabla\,\bm\o+\bm\s\wedge\bm \o+\bm\o\wedge\bm\s,
\ee
and $\bm\o$ can have spatial indices. 

Finally, the last piece which depends on the type of the non-relativistic limit is
\ba
S_\text{type\ A}\eq\frac {\kappa } {2\pi}
\int
\tr\left[-\bm\varsigma\wedge\bm \vartheta\wedge\bm \vartheta
+2\,\e_{ij}\,\bm\vartheta\wedge\left(\bm\pi^i\wedge\O^j
 +\bm\theta^i\wedge E^j+ \bm  \theta^i \wedge \bm \pi^j
 \right)\right],\\
S_\text{type\ B}\eq \frac {\kappa } {2\pi}
\int\tr\Big[
\L\,\bm\varsigma\wedge\bm \varpi\wedge\bm \varpi
+
\epsilon_{ij}\,\bm\varpi\wedge\left(
2\,\bm\theta^i\wedge\O^j
 -2\,\L\,\bm\pi^i\wedge E^j+\bm  \theta^i \wedge \bm \theta^j
 -\L\,\bm  \pi^i \wedge \bm \pi^j\right)
 \Big].\nn
 \ea

\section{Conclusion}
\label{sec: discussion}

In this work, we have studied the non-relativistic limits of three-dimensional colored gravity in terms of generalized \.In\"on\"u-Wigner contractions. 
Let us 
conclude our work with a brief summary of what have been done.
In Section \ref{sec-non-relativistic}, we revisited the three-dimensional non-relativistic isometry algebras and their bilinear forms. 
We reviewed various central extensions of the Galilei and Newton-Hooke algebras. 
Then we discussed their chiral decompositions, how they arise from the contractions of relativistic algebras, 
and the issue of invariant bilinear forms. 
In Section \ref{sec-non-relativistic-colored-alg}, 
we explained the obstruction in the chiral decompositions of non-relativistic colored gravity 
and constructed the consistent non-relativistic colored gravity algebras using  generalized \.In\"on\"u-Wigner contractions. 
In Section \ref{sec-non-relativistic-colored-action},  
we derived the non-relativistic colored gravity actions
in the Chern-Simons formulation using the non-relativistic algebras 
constructed in Section \ref{sec-non-relativistic-colored-alg}.

\acknowledgments

We thank Karapet Mkrtchyan for useful discussions in the early stages of the work. 
WL wants to thank Yang Lei for useful discussions. 
This work is supported by a grant from Kyung Hee University in 2017 (KHU-20171742).

\appendix

\section{Associative structure of relativistic symmetry}

If  we identify the identities $\cI$ and $\tilde \cI$ with $\cZ$ and $\tilde \cZ$, the associative products of $gl(2)\oplus gl(2)$ generators are
\be
\cJ_a\, \cJ_b=-\frac 1 \L \cP_a\, \cP_b=\eta_{ab}\,\cZ_\cJ+\epsilon_{ab}{}^c\, \cJ_c,
\quad
\cJ_a\, \cP_b=\eta_{ab}\,\cZ_\cP+\epsilon_{ab}{}^c\, \cP_c,
\ee
\be
\cZ_\cJ\, \cZ_\cJ=\cZ_\cJ,\quad
\cZ_\cJ\, \cZ_\cP=\cZ_\cP\, \cZ_\cJ=\cZ_\cP,\quad
\cZ_\cP\, \cZ_\cP=-\L\, \cZ_\cJ,
\ee
\be
\cJ_a\, \cZ_\cJ=\cZ_\cJ\, \cJ_a =\cJ_a,\quad
\cJ_a\, \cZ_\cP=\cZ_\cP\, \cJ_a =\cP_a,
\ee
\be
\cP_a\, \cZ_\cJ=\cZ_\cJ\, \cP_a =\cP_a,\quad
\cP_a\, \cZ_\cP=\cZ_\cP\, \cP_a =-\L\,\cJ_a.
\ee
This associative extension of the relativistic isometry algebra can be realized as a tensor product,
\be
\cJ_a=\cL_a\otimes I,\quad
\cP_a=\cL_a\otimes \psi,\quad
Z_\cJ=\cI\otimes I,\quad
Z_\cP=\cI\otimes \psi,\quad
\ee
with
\be
\cL_a\,\cL_b=\eta_{ab}\,\cI+\e_{ab}{}^c\,\cL_c,
\quad
I^2=I,\quad
\psi^2=-\L\, I,\quad
I\,\psi =\psi\, I=\psi.
\ee
The chiral decompositions of the relativistic and the non-relativistic algebras are due to the existence of projectors,
\be
P_L=\frac 1 2 \left(I+ \frac 1{\sqrt{- \L}} \psi\right),\quad
P_R=\frac 1 2 \left(I- \frac 1{\sqrt {-\L}} \psi\right),
\ee
with
\be
P_L^2=P_L,\quad P_R^2=P_R,\quad P_L\,P_R=P_R\,P_L=0.
\ee

\section{Contraction at action level}
\label{contraction-action}

The one-form gauge field \eqref{one-form-A} or \eqref{one-form-B} of the non-relativistic colored gravities should match with the relativistic one,
\ba
\mathcal A&=&(A\,\cZ+\tilde A\,\tilde \cZ
+\o^a\,\cJ_a+e^a\,\cP_a)\otimes I
\nn&&
+(A_I\,\cZ+\tilde {A_I}\,\tilde \cZ
+ \tau_I^a \,\cJ_a  + \chi_I^a\,\cP_a
 )\otimes \bm T^I.
\ea
In this way, we are able to identify the relations between
relativistic and  non-relativistic fields. 
The singlet part is
\be
j=\frac 1 4 (A+\tilde A)+\frac 1 2 \o^0
,\quad
s=\frac {c^2} 2 (A+\tilde A)- {c^2}\, \o^0,
\ee
\be
h=\frac 1 {4\sqrt{-\L}} (A-\tilde A)+\frac 1 2 e^0
,\quad
m=\frac {c^2} {2\sqrt{-\L}} (A-\tilde A)-{c^2}\, e^0,
\ee
\be
\O^i=c\,\o^i,\quad
E^i=c\,e^i.
\ee
The non-relativistic $su(N)$ gauge fields are
\be
\s_I=\frac 1 2(A_I+\tilde A_I),\qquad
\varsigma_I=\frac {c^2} {2\sqrt{-\L}}(A_I-\tilde A_I),
\ee
and the spatial components of the matter fields are
\be
\theta_I^i=c\,\t_I^i,\qquad 
\pi_I^i=c\,\chi_I^i.
\ee
Finally, the temporal components of the matter fields are defined as
\be
\text{Type A}\quad : \quad
\vartheta_I=\tau_I^0,\qquad 
\varpi_I=c^2\,\chi_I^0,
\ee
\be
\text{Type B}\quad : \quad
\vartheta_I=c^2\,\tau_I^0,\qquad 
\varpi_I=\chi_I^0.
\ee
Substituting the relativistic fields with the non-relativistic ones, 
we can derive the non-relativistic actions in Section \ref{explicit-action} 
from the relativistic colored-gravity action, 
by multiplying the action with $c^2$ and taking the limit $c\rightarrow \infty$.

\section{Other contractions of colored gravity algebra}
\label{sec: case01}

In Section \ref{non-rel-alg},
we mentioned two other cases
of non-relativistic limits of the colored gravity algebra.
In the first case, we treat all the non-singlet generators 
as the level-one elements,
whereas in the second case 
we have additional level-zero and level-two elements,
 \eqref{ZZ0} and \eqref{ZZ2}.

 In the first case, the non-relativistic colored generators are defined as
\be
\bm C^I=\frac 1 {c}\,\cZ_\cJ\otimes \bm \cT^I,\quad
\bm D^I=\frac 1 {c}\,\cZ_\cP\otimes \bm \cT^I,
\ee
\be
\bm J^I=\frac 1 {c} \cJ_0\otimes \bm \cT^I,\quad
\bm H^I=\frac 1 {c}\cP_0\otimes \bm \cT^I,
\ee
\be 
\bm G_i^I=\frac 1 {c} \cJ_i\otimes \bm \cT^I,\quad
\bm P_i^I=\frac 1 {c}\cP_i\otimes \bm \cT^I.
\ee
In the non-relativistic limit, 
the non-zero commutators involving colored generators are
\be
[J,\,\bm G_i^I]
=-\frac 1 \L [H,\,\bm P_i^I]
=2\,\epsilon_{ij}\,\bm G_j^I,
\,\quad
[J,\,\bm P_i^I]
=[H,\,\bm G_i^I]=2\,\epsilon_{ij}\,\bm P_j^I,
\ee
\be
[\bm G_i^I,\,
\bm G_j^J]
=-\frac 1 \L [ \bm P_i^I,\,
 \bm P_j^J]
=
\frac 2 {N}\,\epsilon_{ij}\,\d^{IJ}S,
\quad
[\bm G_i^I,\,
\bm P_j^J]
=
\frac 2 {N}\,\epsilon_{ij}\,\d^{IJ}M.
\ee
The relations between relativistic and non-relativistic fields are
\be
 \s_I=\frac c 2\,( A_I+\tilde{  A_I}),\qquad
 \varsigma_I=\frac c {2\sqrt{-\L}}\,( A_I-\tilde { A}_I),
\ee
\be
 \vartheta_I=c\, \tau_I^0,\qquad 
 \varpi_I=c\, \chi_I^0,\qquad
\theta_I^i=c\, \t_I^i,\qquad 
 \pi_I^i=c\, \chi_I^i.
\ee
The non-relativistic action is again given as  \eqref{Action 3}.
$S_\text{gravity}$ is the same as \eqref{gra-action},
whereas $S_\text{gauge}$ and $S_\text{matter}$ are given by
\be
S_\text{gauge}
=
\frac {\kappa } {2\pi}
\int
 \text{\tr }\left(
\bm  \varsigma\wedge d\bm \s
\right),
\ee
\be
S_\text{matter}=
\frac {\kappa } {2\pi}
\int \tr\left[
 -\bm \vartheta\wedge d\bm  \varpi
 +\bm \theta^i\wedge \nabla\,\bm \pi_i
+\epsilon_{ij}\,h\wedge(\bm \theta^i\wedge\bm \theta^j-\L\, \bm\pi^i \wedge \bm\pi^j)
\right]. 
\ee

\bigskip

In the second case,
the level-zero and level-two sectors contain $\cZ_\cJ\otimes\bm \cT^I$ and
$\cZ_\cP\otimes\bm\cT^I$, respectively. 
The non-relativistic generators are defined as
\be
\bm C^I= \cZ_\cJ\otimes \bm \cT^I,\quad
\bm D^I=\frac 1 {c^2} \cZ_\cP \otimes \bm \cT^I,
\ee
\be
\bm J^I=\frac 1 {c} \cJ_0\otimes \bm \cT^I,\quad 
\bm H^I=\frac 1 {c}\cP_0\otimes \bm \cT^I.
\ee
\be
\bm G_i^I=\frac 1 {c} \cJ_i\otimes \bm \cT^I,\quad
\bm P_i^I=\frac 1 {c}\cP_i\otimes \bm \cT^I 
.
\ee

The non-zero Lie brackets involving the colored generators are
\be
[J,\,\bm G_i^I]
=-\frac 1 \L [H,\,\bm P_i^I]
=2\,\epsilon_{ij}\,\bm G_j^I,
\,\quad
[J,\,\bm P_i^I]
=[H,\,\bm G_i^I]=2\,\epsilon_{ij}\,\bm P_j^I,
\ee
\be
[\bm C^I,\, \bm C^J]=2\,i\, f^{IJ}{}_K\, \bm  C^K,
\quad
[\bm C^I,\,\bm D^J]=2\,i\, f^{IJ}{}_K\, \bm D^K,
\ee
\be
[\bm C^I,\,\bm J^J]=2\,i\, f^{IJ}{}_K\, \bm J^K,
\quad
[\bm C^I,\, \bm H^J]=2\,i\, f^{IJ}{}_K\,  \bm H^K,
\ee
\be
[\bm C^I,\,\bm G_i^J]=2\,i\, f^{IJ}{}_K\, \bm G_i^K,
\quad
[\bm C^I,\, \bm P_i^J]=2\,i\, f^{IJ}{}_K\,  \bm P_i^K,
\ee
\be
[\bm J^I,\, \bm H^J]
=
-2\,i\,f^{IJ}{}_K\,\bm D^K,
\ee
\be
[\bm G_i^I,\,
\bm G_j^J]
=
-\frac 1 \L [\bm P_i^I,\,
\bm P_j^J]
=
\frac 2 {N}\epsilon_{ij}\,\d^{IJ}S,
\ee
\be
[\bm G_i^I,\, \bm P_j^J]
=
\frac 2 {N}\epsilon_{ij}\,\d^{IJ}M
+
2\,i\,\d_{ij}\,f^{IJ}{}_K\,\bm D^K.
\ee

The non-relativistic fields are related to the relativistic fields by
\be
\s_I=\frac 1 2(A_I+\tilde A_I),\qquad
\varsigma_I=\frac {c^2} {2\sqrt{-\L}}(A_I-\tilde A_I),
\ee
\be
\vartheta_I=c\,\tau_I^0,\qquad 
\varpi_I=c\,\chi_I^0,\qquad
\theta_I^i=c\,\t_I^i,\qquad 
\pi_I^i=c\,\chi_I^i.
\ee
Again, the action is given by \eqref{Action 3},
where $S_\text{gravity}$
and $S_\text{gauge}$ correspond to \eqref{gra-action} and \eqref{S gauge},
while 
$S_\text{matter}$ coincides with $S_\text{matter,0}$ in \eqref{S matter 0}.

\end{document}